\documentclass{article}
\usepackage[utf8]{inputenc}
\usepackage[affil-it]{authblk}
\usepackage{blindtext}
\title{Performance Analysis of FEM Solvers on Practical Electromagnetic Problems}

\author[1]{\small Gergely Máté Kiss}
\author[2]{\small Jan Kaska}
\author[3]{\small Roberto André Henrique de Oliveira }
\author[4,*]{\small Olena Rubanenko}
\author[5]{\small Balázs Tóth}

\affil[1]{\footnotesize AVL List GmbH, {Graz}, {Austria},}
\affil[2]{\footnotesize Department of Theory of Electrical Engineering, University of West Bohemia, Pilsen, Czech Republic}
\affil[3]{\footnotesize CT/UNL - New University of Lisbon, Caparica, Portugal}
\affil[4]{\footnotesize RICE, University of West Bohemia, Pilsen, Czech Republic}
\affil[5]{\footnotesize Institute of Applied Mechanics, University of Miskolc, Hungary}

\affil[*]{Corresponding author: Olena Rubanenko, rubanenk@rice.cz}

\date{\today}

\usepackage{cite}
\usepackage{graphicx}
\usepackage{amsmath}
\usepackage{amssymb}
\usepackage{gensymb}
\usepackage{hyperref}
\begin{document}

\maketitle

\begin{abstract}
     The paper presents a comparative analysis of different commercial and academic software. The comparison aims to examine how the integrated adaptive grid refinement methodologies can deal with challenging, electromagnetic-field related problems. For this comparison, two benchmark problems were examined in the paper. The first example is a solution of an L-shape domain like test problem, which has a singularity at a certain point in the geometry. The second problem is an induction heated aluminum rod, which accurate solution needs to solve a non-linear, coupled physical fields. The accurate solution of this problem requires applying adaptive mesh generation strategies or applying a very fine mesh in the electromagnetic domain, which can significantly increase the computational complexity. The results show that the fully-hp adaptive meshing strategies, which are integrated into Agros-suite, can significantly reduce the task's computational complexity compared to the automatic h-adaptivity, which is part of the examined, popular commercial solvers.  
\end{abstract}

\let\thefootnote\relax\footnotetext{Accepted in Periodica Polytechnica Electrical Engineering and Computer Science, the paper will be published in https://pp.bme.hu/eecs}

\section{Introduction}

The solution of an industrial design problem generally contains many challenging sub-tasks. The accurate modeling of these electrical devices or machine design problems needs an accurate numerical solution of partial differential equations. These modeling equations usually describes the interaction of different physical fields, respecting their mutual analysis and synthesis. Developing a satisfactory and sufficiently reliable model is difficult because of the mutual interaction of different areas \cite{yilmaz_fe_comparison,Trends_moderne_electrical_machine_design,trafoevo,team,orosz2020fem}. 
Most of the advanced FEM techniques are suitable for solving partial differential equations. However, a design task usually means an optimization procedure, where the design parameters of an electrical machine should be estimated. Due to a large number of the design variables, the solution space is large. These coupled FEM models are generally computationally expensive \cite{tenne2010computational,team_benchmark,orosz2017performance}. Practical design problems generally have to deal with manufacturing tolerances. The applied materials contain non-linearities, inhomogenous, or anisotropic material properties. These should also be considered during the optimization \cite{dibarba,dolevzel2014multi,di2014multiphysics,rassolkin2020life}.

Adaptive grid refinement has a crucial role in the improvements of the solution of partial differential equations (PDE), which is the most computationally-intensive part in the case of a wide range of engineering design tasks \cite{mitchell2013collection,plaszewski2013performance}. Many of the applications at the cutting edge of research are extraordinarily challenging. There are a lot of different problems during the solution of different problems \cite{mitchell2014comparison}.
In our work, we are focusing only on the solution of the elliptic partial differential equations, because the solution of the electromagnetic problems generally leads to these kinds of PDEs \cite{mitchell2014comparison}.

Self-adaptive methods have been studied around 40 years now \cite{szabo1979some,babuska1982rates}. 
An algorithm is adaptive if it contains some local a posteriori error estimation capability (energy norm, $H^1$ norm, $L^2$ norm, etc.) and the capability to increase the number of degrees of freedom (DOF) with or without a minimal user-interaction. 
The automatic adaptivity algorithms can be divided into the following groups:  h-adaptivity, p-adaptivity, hp-adaptivity. In a finite element method (FEM), the problem's solution space is discretized into nodes and elements, where the function value is approximated by a basis function, which is usually a polynomial. During an h-adaptivity based refinement, only the mesh is refined, while a p-adaptivity based refinement changes the degree of the approximating polynomial in the given element \cite{mitchell2014comparison}.The h-adaptive methods are quite well understood now, and they are usually implemented in commercial codes \cite{ANSYS,COMSOL,luo2016novel}.
Academic researchers pay more attention on the hp-adaptive techniques. These methods adapt the size of the mesh and the degree of the polynomials, as well. Therefore, the error can converge with an exponential rate. In contrast to the \textit{h}- and \textit{p}-adaptivity methods. It shows the complexity of the hp-adaptivity that many papers showed it's advantages in the 1980s; however, the first practical implementation was published only in the 1990s. Because the local estimators cannot be used in this case to guide the adaptivity, they can provide the information that which elements should be refined, however, they can not indicate which kind of adaptivity should be used (\textit{h} or \textit{p}) on the selected element.
A method for making that determination is called a hp-adaptive strategy \cite{mitchell2014comparison, szabo1979some}.
Many novel and promising hp-adaptive strategies exist in the literature \cite{ainsworth1997aspects}.
It is not clear which ones perform best under different situations or if any of the strategies are good enough to be used as a general-purpose solver.
Most of them are based on well-posed multi-field variational formulations \cite{FueKeiDemTal17,TB16,TB18,TB19}. One of them is the discontinuous Petrov--Galerkin (DPG) method. This method was elaborated on an ultraweak variational formulation. All the derivatives are shifted to the test functions, leading to non-symmetric functional settings, i.e., the trial and test functions do not arise from the same function space. The practical DPG methods apply broken (discontinuous) function spaces to approximate the test functions at the element level. The DPG method is equivalent to a minimum residual method, and, also, it can be verified that the DPG method results in a mixed method, where a certain type of the residual defines an error representation function serving as a local error indicator, guiding the adaptive procedure, see the details, for example in \cite{FueKeiDemTal17,Niemi11}.

However, the performance most of these novel methods are not investigated in this paper. In this article, we present a practical comparison between existing and widely used commercial and open-source applications.
Those tools are preferred, which used to solve electromagnetic problems and have a user-friendly interface. Thus, those academic and open-source projects, which contains high quality codes, libraries, but requires deep knowledge in programming are not considered in this practical comparison (like, FEniCS \cite{FENICS}, FreeFem++ \cite{FreeFem}, GetFEM++ \cite{GetFEM} or GetDP \cite{getdp}). The Ansys, JMAG, and the COMSOL Multiphysics are used for the comparisons from the commercial applications. 

From these tools, only the Agros-suite support fully hp-adaptive strategies \cite{hermes-adaptivity, hermes-hanging-nodes, hermes-runge-kutta} and FEMM has an easy to use interface but it doesn't support automatic adaptivity the mesh can be refined only manually. While JMAG does support adaptive mesh refinement, it does not enable it in problems where Neumann boundary condition is applied. Therefore in case of JMAG, as in FEMM, manual mesh refinement is applied. Agros also supports time adaptivity, multi-meshing technology and can handle the hanging nodes. Time adaptivity can be useful to reduce the computational complexity of transient problems. Multimeshing technology is important in the case of coupled problems \cite{mifune2002fast}, like the second showed induction heating example, where the accurate calculation of the skin effect needs a very fine mesh, however, a rough mesh is enough for the calculation of a temperature field related PDE. Hanging nodes can increase the computational complexity, but the implemented algorithm can allow the mesh refinement around the closest domain around the hanging node, while external parts require no additional refinement \cite{karban2020fem}. 

The paper shows two practical benchamark examples from the field of 2D electromagnetic calculations. These problems solved by various solvers and  the results

\section{Example I}

\subsection{Electrostatic spark gap}

Firstly we will start our investigation with an electrostatic variant of the classical  L-shape domain like test problem \cite{mitchell2013collection}. This problem is good to examine the singular point handling performance of the different solvers. The singular point means a point at the examined geometry, where is not possible to define the normal and the gradient of the solution, because it grows there over all the numerical limits \cite{karban2013advanced}. 

An electrostatic spark gap consists of two conducting electrodes, which are separated by a gap usually filled with a gas insulation, like air or more efficient insulators, like SF6 gas \cite{tsuboi1994computation,said1997case}. This device is designed to allow an electric spark to pass between the conductors. When the potential difference between the conductors exceeds the breakdown voltage of the gas within the gap, a spark forms, ionizing the gas and drastically reducing its electrical resistance. Then an electric current flows in the ioinized gas of the gap, which reduces the overcurrent in the protected part of the electrical circuit. 

\begin{figure}[h!]
\centering
\includegraphics[scale=0.7]{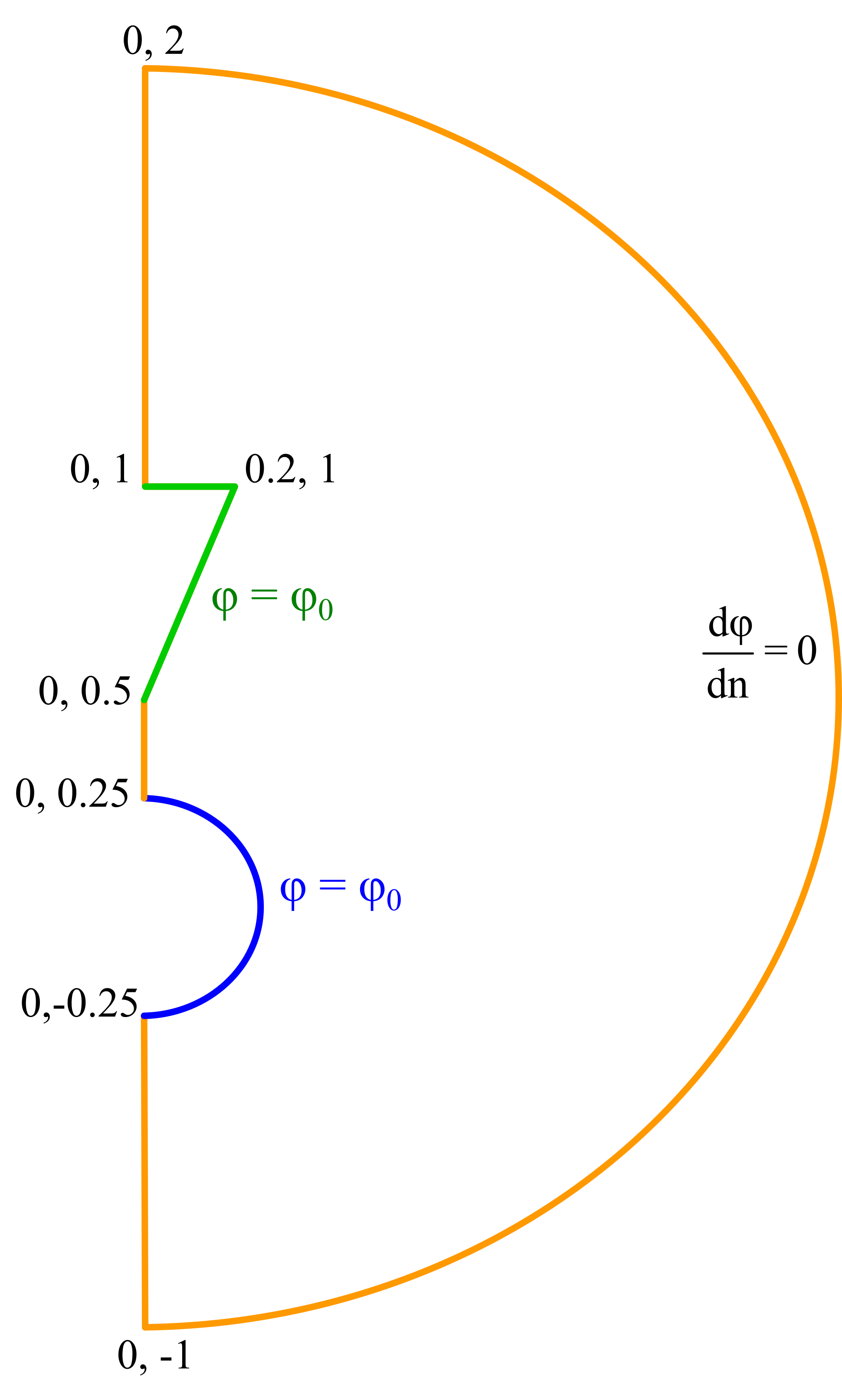}
\caption{The geometry of the Electrostatic test problem, every dimension is given in [m]. The applied voltages are $\phi_1 = 1000 V$ and $\phi_0 = 0 V$. }
\label{fig:universe}
\end{figure}

The examined 2D axisymmetric geometry and the applied boundary conditions are plotted in Fig.\ref{fig:universe}. Where, homogeneous Dirichlet-type boundary conditions are set on the two lateral surfaces, which meets at the re-entrant corner. Neumann-type BCs are prescribed on the remaining boundaries. The derivatives of the solution has a singularity at the origin, this singularity comes from the non-convexity of the examined region. This problem is previously analyzed by different authors \cite{karban2013advanced,femmimp}. However, during these different comparisons the shape and the position of the outer boundary layer was selected differently (rectangular or circular shaped outer boundaries). However, due to the numerical error of the 2D axysimmetrc methods, the electrostatic energy increases if we are integrating it on a larger airgap:

\begin{equation}
    w_\mathrm{e} = \frac{1}{2} \vec{E} \vec{D}, 
\end{equation}
\begin{equation}
  W_\mathrm{e} = \int_0^{2 \pi} \int_0^r \int_0^z  w_\mathrm{e} r \mathrm{d} z \mathrm{d} r \mathrm{d} \alpha.
\end{equation}
Where  $w_\mathrm{e}$ is the density of the electrostatic energy, $E$ is the electric field strength vector and the $D$ is the electric displacement field, $W_\mathrm{e}$ is the energy of the electrostatic field. 
We can approximate the error of this integral if we consider the layout from far away as a dipole. In this case the electric field changes $E \approx 1/r^2$, so the electrostatic energy error $W_e \approx 1/r^4$, while the error of the solvers in the range of $W_e \approx 10^{-6}$, so it is important to use exactly the same geometry during the analyses and compare the current version of these solvers again.

\subsection{Results and analyses}

\begin{figure}[h!]
\centering
\includegraphics[width=0.7\linewidth]{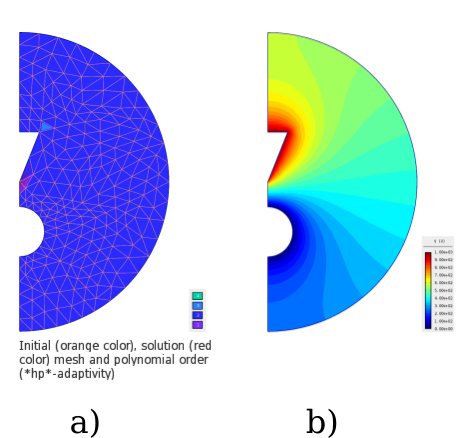}
\caption{The generated hp-adaptive mesh in Agros2D (a) and the distribution of the electric potential (b) in the airgap.}
\label{fig:we}
\end{figure}

The problem was solved by ANSYS, COMSOL, FEMM 4.2 (version 21Apr2019) and Agros2D. We would like to use another popular FEM softwares, however JMAG and MAGNET doesn't contains a 2D axysymmetric electromagnetic solver. The commercial codes contains automatic h-adaptivity, while in case of Agros2D we used the full automatic hp-adaptivity with 2nd order elements and curvilinear elements for the better approximation of the outer boundary (Fig. \ref{fig:we}). In case of FEMM, there is no option for an automatic hp-adaptivity, however we continously refined the mesh manually, doubling the resolution (max element size is set to x*0.5 in all domains using the F3 command in FEMM) with the built-in function of the mesh generator.

There were the convergence of the following two quantities examined during the calculations:

\begin{itemize}
    \item electrostatic energy in the air gap. 
    \item electric field stress in a point, very close to the singularity: r = 0.01 m, z = 0.5 m. 
\end{itemize}

\begin{figure}[h!]
\centering
\includegraphics[width=0.95\linewidth]{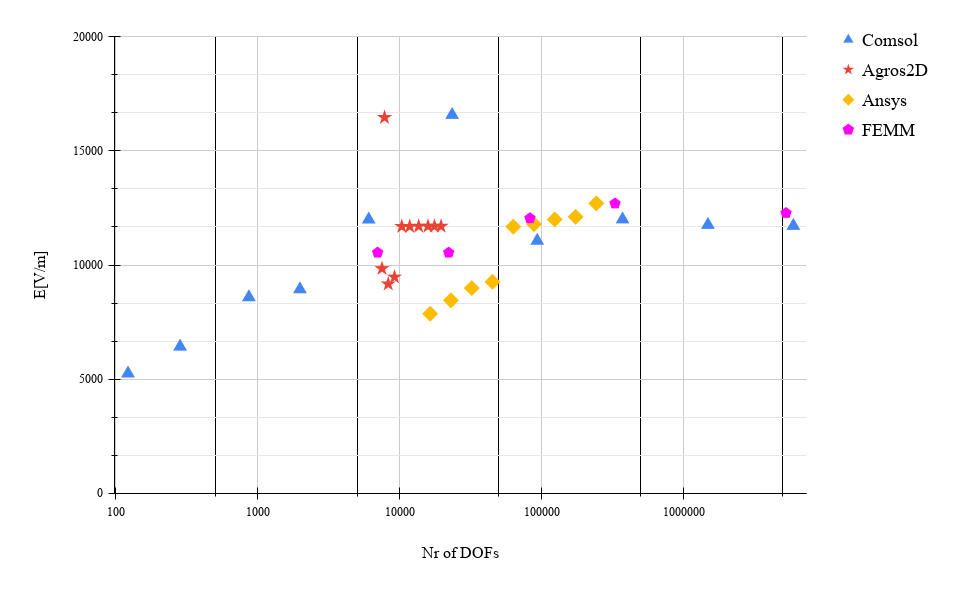}
\caption{Convergence of the different solvers in case of the electric field stress in point r = 0.01 m , z = 0.5 m. }
\label{fig:e}
\end{figure}

Fig. \ref{fig:e} summarizes the numerical results of the electrostatic energy calculation with the different solvers in the function of the DOFs. It can be seen from the results that the electrostatic energy converges in all cases to 7.63e-6 J, except in the case of the Ansys solution, which differs about 3-4\% from the other solutions. Agros2D reaches this final number with using only 1780 elements for the calculations. Therefore, Agros2D needs more than eight times less elements to reach the same numerical precision than COMSOL Multiphysics (10391) or more than one hundred times less elements than the open-source FEMM. The convergence behaviour of the ANSYS solution seems similar like COMSOL Multiphysics.

In the second case (Fig \ref{fig:e}) the local convergence is examined near the point  r = 0.01 m, z = 0.5 m. It can be seen that the local convergence is more slower for all of the examined solvers. The results converges to 1.17e4 kV/m, the convergence of the Agros2D is the fastest in this case, as well. It reaches this value with the usage of 10345 elements, while COMSOL needs 93107 elements converge to the similar result.

\section{Example II}

\subsection{Coupled Field analysis}

The second example demonstrates a typical induction heating application. The process of induction heating is commonly used in the manufacturing industry.
Use cases spread from surface hardening through annealing to shrink fit assembly. Induction heating is a fairly simple, clean, and quick way to heat up workpieces \cite{rudnev2017handbook}. The only requirement being is that the part to be heated shall produce substantial losses if subjected to an alternating magnetic field. The alternating magnetic field is mostly produced by an induction heating coil supplied by high current and high-frequency power supply. Heating coils are specifically designed for the application as their geometry largely depends on the part to be heated. Their heating characteristic has to fulfill a number of design requirements:
\begin{itemize}
 \item{they might have to heat a specific area of the manufactured part only,}
 \item{material dependent local maximal temperatures must not be exceeded,}
 \item{an even temperature profile might be required, which can be difficult to achieve in case of complex parts,}
 \item{process time must meet demands of production, etc.}
\end{itemize}

The designed application should be insensitive to the manufacturing tolerances and the positioning errors. Moreover, the material properties can be anisotropic or inhomogenous, and the mechanical and electromagnetic properties of the workpiece can be non-linear functions of the temperature of the workpiece. For example, during the hardening of a steel workpiece, its magnetic permeability decreases to $\mu_0$ till it reaches the Curie-temperature \cite{dibarba}.
Designing a robust induction process is a challenging task in the industry. In this case, the heating coil has to be designed together with temperature control of the process \cite{artap_paper, induction_brazing_1, induction_brazing_2,panek2020comparison,dolezel2010accurate,karban2020fem}.

\subsection{Eddy loss analysis}
\label{sec:eddy_loss_section}

In general, an induction heating problem is often starting as a 3D model and then simplified to a 2D axisymmetric problem for the optimization rounds. The simplification may often be done due to the cylindrical symmetric nature of heating coils. Induction heating coil themselves are in most cases tubular, and are actively cooled with liquid. Therefore temperature dependent behavior of the coil material can be neglected. Material properties of the workpiece however are temperature-dependent and non-linear, especially in the case of ferromagnetic materials. To accurately model the process of induction heating, a coupled analysis is often required. A coupled analysis works by linking the electromagnetic and thermal domains together, solving the problem quasi simultaneously in both domains. The coupling may operate in way that as a first step an electromagnetic problems are solved in the frequency domain, providing the Joule loss as a result. The result is then picked up by the coupled transient thermal model, which computes the temperature distribution using the losses. In the next step the electromagnetic solver calculates the losses using the changed material properties due to the heating. This iterative process is continued until an equilibrium state is reached, or the material was at target temperature. In such a case it is possible to use temperature-dependent material data for improved accuracy of results. This, of course, requires very precise characterization of the materials used, including temperature dependent magnetic permeability, electric conductivity, heat capacity, heat conductivity etc. as seen in \cite{dibarba}.

Another crucial point of modeling is choosing a mesh with an adequate resolution. Induction heating uses high frequency (usually f $>$ 1 [kHz]) magnetic fields in order to achieve the desired temperatures. As a result, most of the losses would appear near the surface of parts due to skin effect.
The penetration depth of the magnetic field generated by the coil or the skin depth can be expressed as follows:

\begin{equation}
    \delta ={\sqrt {{2\rho } \over {\omega \mu }}}{\sqrt {{\sqrt {1+\left({\rho \omega \epsilon }\right)^{2}}}+\rho \omega \epsilon }}
    \label{eq:long_skin_depth}
\end{equation}
where

\begin{itemize}
    \item{ $\rho$ : resistivity of the conductor,}
    \item{$\omega$ : angular frequency of current = $2\pi\times frequency$, }
    \item{$\mu$ : permeability of the conductor $\mu_{0} $,}
    \item{$\mu_{r}$ : relative magnetic permeability of the conductor,}
    \item{$\mu_{0}$ : the permeability of free space,}
    \item{$\epsilon$ : permittivity of the conductor $\epsilon_{0} \epsilon_{r}$,}
    \item{$\epsilon_{r}$ : relative permittivity of the conductor,}
    \item{$\epsilon_{0}$ the permittivity of free space.}
\end{itemize}

At frequencies used by induction heating and the equation \ref{eq:long_skin_depth} can be simplified to the following from:

\begin{equation}
    \delta ={\sqrt {{2\rho } \over {\omega \mu }}}
    \label{eq:skin_depth}
\end{equation}

Equation \ref{eq:skin_depth} shows that the resulting skin depth, beside the frequency, is highly influenced by the material's resistivity and permeability. Meaning that the problem is more complicated in case of soft magnetic materials as it may vary locally along the surface due to uneven magnetic field strength.

\begin{figure}[h!]
\centering
\includegraphics[width=0.95\linewidth]{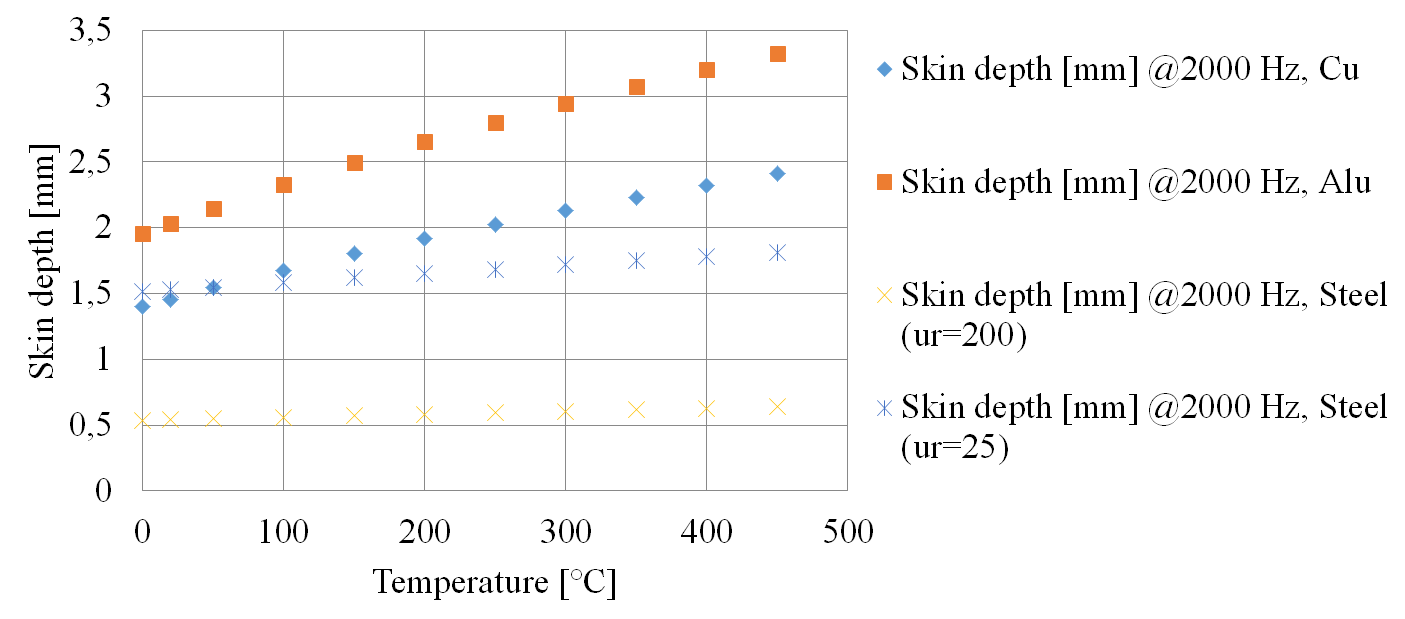}
\caption{Variation of skin depth depending on temperature in case of different materials}
\label{fig:eddy_CuAlFe}
\end{figure}

\begin{figure}[h!]
\centering
\includegraphics[width=0.95\linewidth]{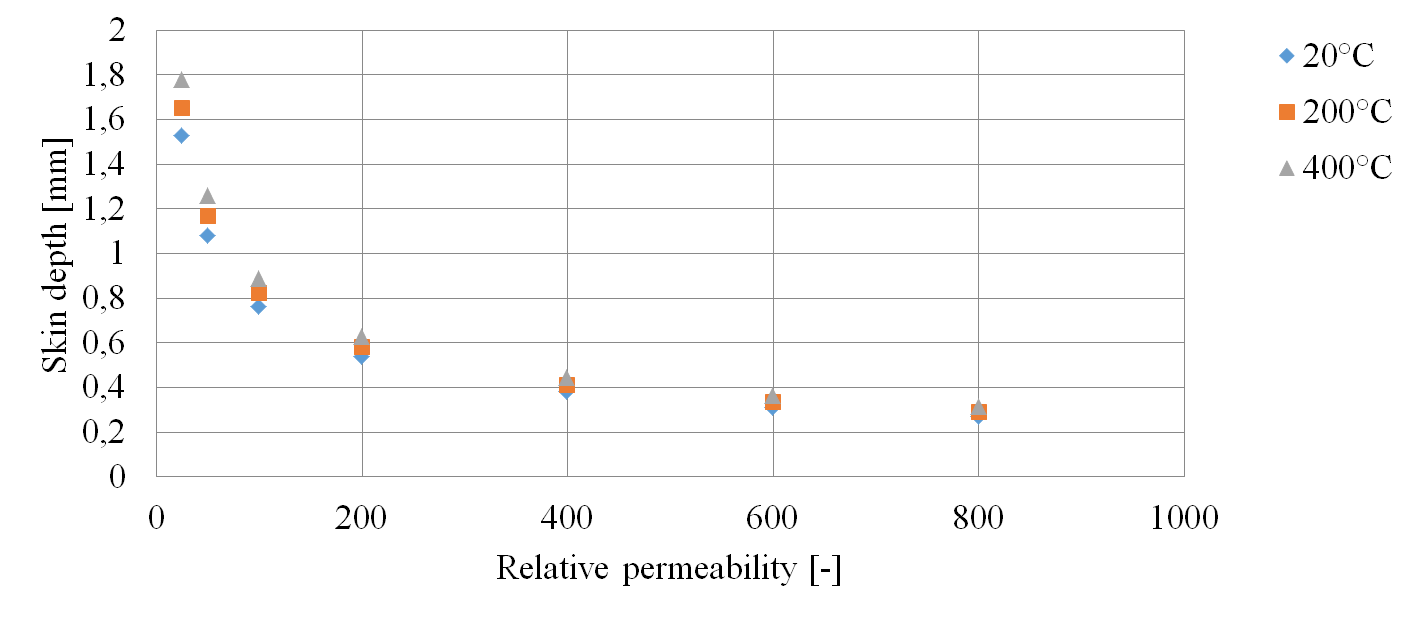}
\caption{Variation of skin depth in case of ferromagnetic materials (eg. steel)}
\label{fig:eddy_FeUrT}
\end{figure}

In order to demonstrate the meaning of Equation \ref{eq:skin_depth} Figure \ref{fig:eddy_CuAlFe} shows how skin depth is changing depending on the temperature in case of different materials and same field frequency. It can be observed how generally good conductors like copper and aluminium have a high sensitivity to temperature changes, due their conductivity's high temperature dependency. Meanwhile a steel material appears to be significantly less sensitive to temperature change. This is of course only true as long as the permeability remains constant over the investigated temperature range. Figure \ref{fig:eddy_FeUrT} shows how sensitive skin depths of ferromagnetic materials can be to variation of relative permeability. The magnetic permeability of a material is in general a function of both H [A/m] magnetic field strength and T [K] temperature up to its specific Curie-temperature, where it approaches $\mu_{0} $. Knowledge of the $\mu_{0}(H,T) $ function is extremely useful if modeling accuracy is the highest priority.

The resulting skin depth for an aluminium part at room temperature and 2 [kHz] frequency is at 2 [mm]. Which means, depending on the size of the part, only a small portion of the model is actually significant for the results. Thus the part should have a fine enough mesh in the area where eddy currents are forming, but only a coarse mesh everywhere else. The difficulty is, that the skin depth cannot be accurately predicted in advance, before the simulation, as the magnetic field distribution is unknown. For example the same aluminium part will have a 3 [mm] skin depth once it reaches 300 [\degree C] temperature. According to Figure \ref{fig:eddy_FeUrT} steel's skin depth can vary in a large range, even at the same temperature, depending on the magnetic field strength which influences permeability directly. Precautiously built models therefore apply fine mesh density over the entire model, causing a large number of elements and consequently quite long solution time. Adaptive meshing methods can however refine the mesh at the most critical areas, at the expense of some computation time as well due to refinement steps.

\subsection{Model description}

\begin{figure}[h!]
\centering
\includegraphics[width=0.45\linewidth]{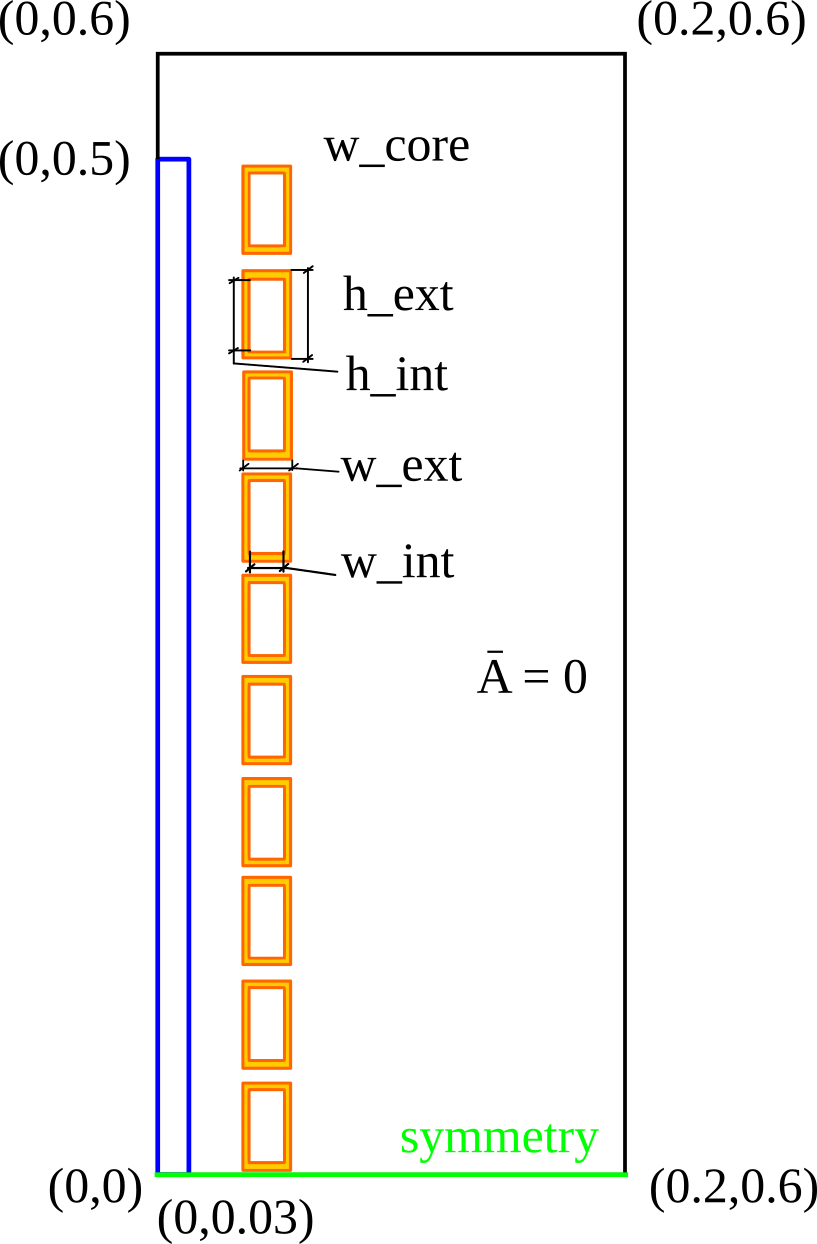}
\caption{Geometry used for eddy current loss analysis}
\label{fig:eddy_geo}
\end{figure}

The scope of the current paper extends to the performance analysis of different solvers only, the absolute modeling accuracy of induction heating is not the goal this time. Therefore the model presented in this paper shows the induction heating application in its simplest form, consisting purely of the coil and a metallic work piece that is to be heated through eddy current loss. Figure \ref{fig:eddy_geo} demonstrates the geometry used for this example. Table \ref{tab:eddytable1} lists the used material parameters, shows additional model dimensions and the applied current.
Both parts are symmetrical to enable use of model size reduction by applying a symmetry boundary condition. As a further simplification the coil is assumed to be free of eddy currents, thus only the phenomena taking place in the aluminium cylinder is considered.

The method of mesh refinement is by using adaptive techniques where applicable (cases Agros2D, Ansys and COMSOL Multiphysics) and reduction of global element size where adaptive techniques are not supported (cases FEMM and JMAG).

\begin{table}[]
\centering
\caption{Parameters of eddy loss model}
\label{tab:eddytable1}
\begin{tabular}{lll}
\hline
\multicolumn{3}{l}{Model parameters}                   \\ \hline
Conductivity of copper             & 0       & MS/m    \\
Conductivity of aluminium          & 30.6327 & MS/m    \\
Relative permeability of all parts & 1       & {[}-{]} \\
Coil current (rms)                 & 3500    & A       \\
Current frequency                  & 2000    & Hz      \\
w\_ext                             & 0.02    & m       \\
w\_int                             & 0.014   & m       \\
h\_ext                             & 0.04    & m       \\
h\_int                             & 0.034   & m       \\ \hline
\end{tabular}
\end{table}


\subsection{Eddy loss calculation results}

\begin{figure}[h!]
\centering
\includegraphics[width=1.02\linewidth]{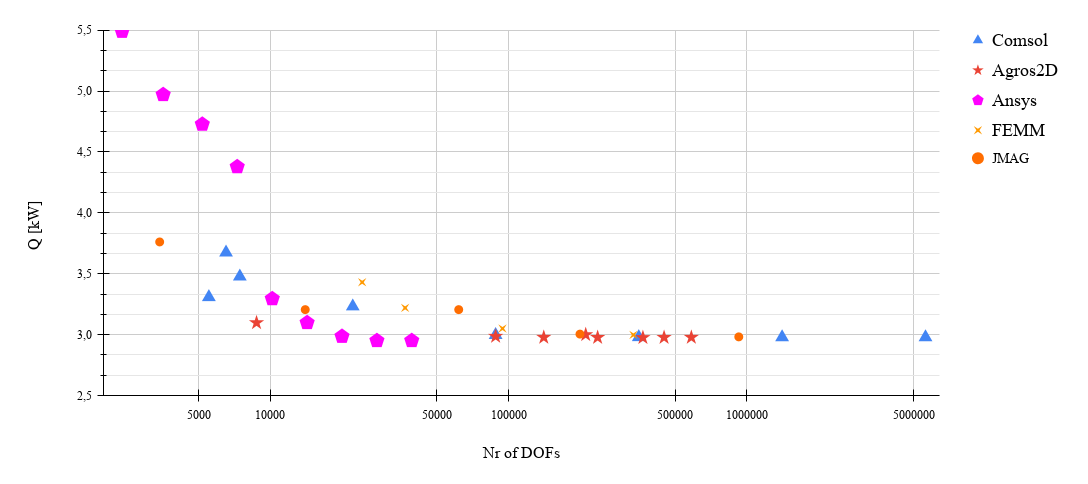}
\caption{Convergence of the different solvers in case of the eddy loss calculation in the aluminum tube. }
\label{fig:eddy}
\end{figure}

The generated eddy loss is calculated by different solvers (ANSYS, COMSOL, JMAG, FEMM, Agros2D) in the volume of the heated aluminum tube. It can be seen from the results that there is no huge difference between the different solvers like in the previous calculation. Agros2D, Ansys and Comsol has the same precision after 100 000 calculations, JMAG and FEMM also has a similar but a bit slower convergence to the resulting 2.98 kW. According to Fig. \ref{fig:eddy} solvers of ANSYS and Agros2D converge the fastest, however the difference is not as significant, as in Example I.

\begin{figure}[h!]
\centering
\includegraphics[width=0.95\linewidth]{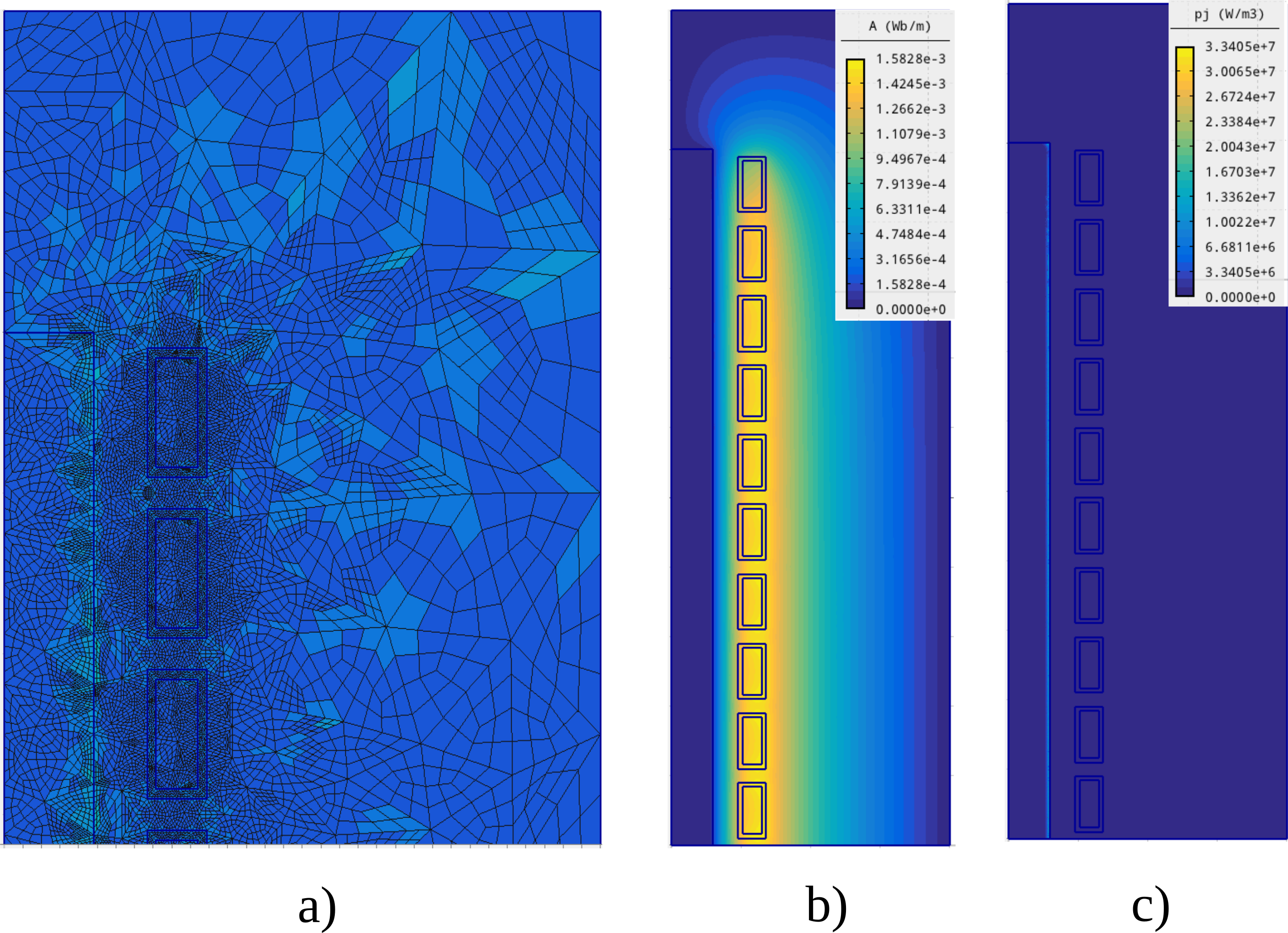}
\caption{The picture a plots the generated hp-adaptive mesh by Agros2D on the part of the geometry. Pictur b) plots the calculated flux lines and the magnetic vector potential, while picture c) plots the calculated eddy losses in the aluminum rod.}
\label{fig:lines_flux}
\end{figure}

\subsection{Coupled Field Analysis}
\label{sec:CoupledFieldAnalysis}

The full solution of the example needs not only a magnetic field analysis, but the thermal field should be considered, as well. However, the material properties of the aluminum (heat capacity, density, heat and electrical conductivity) have a strong dependency on the temperature and all have a significant effect on the final temperature distribution of the heating process. Figure \ref{fig:temp} demonstrates this effect, where the red dots illustrates, how the hot-spot temperature of the aluminum rod changes if the temperature dependent parameters are not considered during the induction heating, while the blue dots represents the solution of the non-linear problem by Agros2D. During this calculation, the temperature dependence of the aluminum rod is neglected. The other numerical problem here is coming from the different behaviour of the numerical fields, the first, eddy loss calculation problem needs a very fine mesh close to the edge of the aluminum rod, while the temperature field and the non-linear iteration during the transient heating process can be accurate and much more faster with a coarser mesh. To overcome this problem, Agros uses a multimesh technology for these calculations.

Based on Fig.\ref{fig:eddy_CuAlFe}, it can be concluded that in case of Example II only approximately the upper 2-4 [mm] radial layer of the aluminium part is playing an active role in heat generation from the start till the end of the process. This translates to 6.67-13.3\% of the total area of the part in the axisymmetric model according to dimensions seen in Fig. \ref{fig:eddy_geo}. This proportion may become even smaller in case of higher current frequencies or different materials (eg. steel). It is clear that the skin depth area, where Eddy-currents are present and Joule losses are forming, is going to be the most significant part of the electromagnetic domain. Therefore the skin depth region requires an adequately fine mesh. However prediction of the skin depth in advance of the calculation is not possible due to non-linearity of the material parameters. As we could see in Sec. \ref{sec:eddy_loss_section} variation of permeability has a significant impact on skin depth. Due to the unknown skin depth, moreover the distribution of skin depth: since it may vary spatially along the investigated part, the mesh is chosen to represent a 'worst-case' scenario. The goal is to make sure the model can reflect a realistic heating pattern. As one could expect, this usually results in model with an extremely high element number. Large models, as a rule of thumb, are computationally expensive, especially if nonlinear ferromagnetic materials are used in the magnetic domain. Materials with a nonlinear BH-curve in frequency domain analysis are usually handled by different numerical techniques in the different solvers solver-to-solver. The nowadays used processes usually involves linearization around specific points in the curve as no saturation-related harmonics can be part of the solution.

Opening up the scope to the thermal part of the coupled analysis we can see that the non-linearity of the induction heating problem is present even in case of magnetically linear materials, where relative permeability does not change, like aluminum. Figure \ref{fig:eddy_CuAlFe} show how skin depth will vary in a nonlinear way depending on temperature. Based on this we can conclude that even magnetically linear materials may have a spatially non-uniform skin depth distribution, making coupled analysis an essential method for similar applications. Special attention could be paid to the fact, that the thermal model requires significantly lower mesh resolution compared to the electromagnetic domain. However the resolution shall still be high enough to accurately represent the heat sources (i.e. the Joule loss distribution itself) in the model. Therefore accurately mapping the losses calculated by the electromagnetic solver into the thermal model, and then transferring the material temperature changes back into the electromagnetic domain are the most crucial steps in coupled analysis. 

In practical use coupled analysis if often applied in order to optimize an induction heating coil shape or the geometry of the heated part for a certain heating application. The process of optimization requires many iterative simulations of the same model with slight changes, in order to reach the optimization target. It is easy to conclude that large, precise, over-meshed models might result in a computationally expensive optimization process, either meaning slowly appearing results or high utilization of a cluster, likely with increased costs. However models which lack the overall accuracy, might not worth being used for optimization as the results could be sub-optimal.
Utilization of adaptive solvers, that take into account the required precision for each step in both electromagnetic and thermal domain and continuously keeping the mesh fit for the purpose, might resolve this conflict.


\section{Conclusions}





The paper presented a practical performance comparison of selected commercial and open-source tools. The goal of the analysis is to compare some FEM tools widely used in the industry with similar, open-source FEM tools, which have a user-friendly interface. The subject of the analysis is the electromagnetic design and show to the potential of the hp-adaptive mesh generation in the electrical design problems. Only the Agros-suite has a fully hp-adaptive mesh generation library from the examined tools. It can be seen that these novel methods can significantly decrease the computational complexity and time in the future. The result of the first example showed that these hp-adaptive algorithms could converge ten times faster than the automatical h-adaptivity, which is built into the commercial codes. There are also minor differences between the commercial tools' converges, but it seems not significant and depends on the selected problem.

\bibliographystyle{ieeetr}
\bibliography{references}

\begin{thebibliography}{10}

\bibitem{yilmaz_fe_comparison}
G.~Y. {Sizov}, D.~M. {Ionel}, and N.~A.~O. {Demerdash}, ``A review of efficient
  fe modeling techniques with applications to pm ac machines,'' in {\em 2011
  IEEE Power and Energy Society General Meeting}, pp.~1--6, July 2011.

\bibitem{Trends_moderne_electrical_machine_design}
G.~{Bramerdorfer}, J.~A. {Tapia}, J.~J. {Pyrhönen}, and A.~{Cavagnino},
  ``Modern electrical machine design optimization: Techniques, trends, and best
  practices,'' {\em IEEE Transactions on Industrial Electronics}, vol.~65,
  pp.~7672--7684, Oct 2018.

\bibitem{trafoevo}
T.~Orosz, ``Evolution and modern approaches of the power transformer cost
  optimization methods,'' {\em Periodica Polytechnica Electrical Engineering
  and Computer Science}, 2019.
\newblock to be published.

\bibitem{team}
P.~Karban, D.~Pánek, T.~Orosz, and I.~Doležel, ``Semi-analytical solution for
  a multi-objective team benchmark problem,'' {\em Periodica Polytechnica
  Electrical Engineering and Computer Science}, 2020.
\newblock In Press.

\bibitem{orosz2020fem}
T.~Orosz, D.~P{\'a}nek, and P.~Karban, ``Fem based preliminary design
  optimization in case of large power transformers,'' {\em Applied Sciences},
  vol.~10, no.~4, p.~1361, 2020.

\bibitem{tenne2010computational}
Y.~Tenne and C.-K. Goh, {\em Computational intelligence in expensive
  optimization problems}, vol.~2.
\newblock Springer Science \& Business Media, 2010.

\bibitem{team_benchmark}
P.~Karban, D.~P{\'a}nek, T.~Orosz, and I.~Dole{\v{z}}el, ``Semi-analytical
  solution for a multi-objective team benchmark problem,'' {\em arXiv preprint
  arXiv:2008.06954}, 2020.

\bibitem{orosz2017performance}
T.~Orosz, B.~Borb{\'e}ly, and Z.~{\'A}. Tamus, ``Performance comparison of
  multi design method and meta-heuristic methods for optimal preliminary design
  of core-form power transformers,'' {\em Periodica Polytechnica Electrical
  Engineering and Computer Science}, vol.~61, no.~1, pp.~69--76, 2017.

\bibitem{dibarba}
P.~Di~Barba, M.~E. Mognaschi, D.~Lowther, F.~Dughiero, M.~Forzan, S.~Lupi, and
  E.~Sieni, ``A benchmark problem of induction heating analysis,'' {\em
  International Journal of Applied Electromagnetics and Mechanics}, vol.~53,
  pp.~1--11, 10 2016.

\bibitem{dolevzel2014multi}
I.~Dole{\v{z}}el, P.~Di~Barba, M.~Forzan, and E.~Sieni, ``Multi-objective
  design of a power inductor: a benchmark problem of inverse induction
  heating,'' {\em COMPEL: The International Journal for Computation and
  Mathematics in Electrical and Electronic Engineering}, 2014.

\bibitem{di2014multiphysics}
P.~Di~Barba, I.~Dolezel, P.~Karban, P.~Kus, F.~Mach, M.~Mognaschi, and
  A.~Savini, ``Multiphysics field analysis and multiobjective design
  optimization: a benchmark problem,'' {\em Inverse Problems in Science and
  Engineering}, vol.~22, no.~7, pp.~1214--1225, 2014.

\bibitem{rassolkin2020life}
A.~Rass{\"o}lkin, A.~Belahcen, A.~Kallaste, T.~Vaimann, D.~V. Lukichev,
  S.~Orlova, H.~Heidari, B.~Asad, and J.~P. Acedo, ``Life cycle analysis of
  electrical motordrive system based on electrical machine type,'' {\em
  Proceedings of the Estonian Academy of Sciences}, vol.~69, no.~2,
  pp.~162--177, 2020.

\bibitem{mitchell2013collection}
W.~F. Mitchell, ``A collection of 2d elliptic problems for testing adaptive
  grid refinement algorithms,'' {\em Applied mathematics and computation},
  vol.~220, pp.~350--364, 2013.

\bibitem{plaszewski2013performance}
P.~P{\l}aszewski and K.~Bana{\'s}, ``Performance analysis of iterative solvers
  of linear equations for hp-adaptive finite element method,'' {\em Procedia
  Computer Science}, vol.~18, pp.~1584--1593, 2013.

\bibitem{mitchell2014comparison}
W.~F. Mitchell and M.~A. McClain, ``A comparison of hp-adaptive strategies for
  elliptic partial differential equations,'' {\em ACM Transactions on
  Mathematical Software (TOMS)}, vol.~41, no.~1, pp.~1--39, 2014.

\bibitem{szabo1979some}
B.~A. Szab{\'o}, ``Some recent developments in finite element analysis,'' {\em
  Computers \& Mathematics with Applications}, vol.~5, no.~2, pp.~99--115,
  1979.

\bibitem{babuska1982rates}
I.~Babuska and B.~Szabo, ``On the rates of convergence of the finite element
  method,'' {\em International Journal for Numerical Methods in Engineering},
  vol.~18, no.~3, pp.~323--341, 1982.

\bibitem{ANSYS}
``{ANSYS}.''

\bibitem{COMSOL}
``{COMSOL Multiphysics}.''

\bibitem{luo2016novel}
G.~Luo, R.~Zhang, Z.~Chen, W.~Tu, S.~Zhang, and R.~Kennel, ``A novel nonlinear
  modeling method for permanent-magnet synchronous motors,'' {\em IEEE
  Transactions on Industrial Electronics}, vol.~63, no.~10, pp.~6490--6498,
  2016.

\bibitem{ainsworth1997aspects}
M.~Ainsworth and B.~Senior, ``Aspects of an adaptive hp-finite element method:
  Adaptive strategy, conforming approximation and efficient solvers,'' {\em
  Computer Methods in Applied Mechanics and Engineering}, vol.~150, no.~1-4,
  pp.~65--87, 1997.

\bibitem{FueKeiDemTal17}
F.~Fuentes, B.~Keith, L.~Demkowicz, and P.~{Le Tallec}, ``Coupled variational
  formulations of linear elasticity and the {DPG} methodology,'' {\em Journal
  of Computational Physics}, vol.~348, pp.~715--731, 2017.

\bibitem{TB16}
B.~T{\'o}th, ``Multi-field dual-mixed variational principles using
  non-symmetric stress field in linear elastodynamics,'' {\em Journal of
  Elasticity}, vol.~122, pp.~113--130, 2016.

\bibitem{TB18}
B.~T{\'o}th, ``Dual and mixed nonsymmetric stress-based variational
  formulations for coupled thermoelastodynamics with second sound effect,''
  {\em Continuum Mechanics and Thermodynamics}, vol.~30, no.~2, pp.~319--345,
  2018.

\bibitem{TB19}
B.~T{\'o}th, ``Hybridized dual-mixed $hp$-finite element model for shells of
  revolution,'' {\em Computers \& Structures}, vol.~218, pp.~123--151, 2019.

\bibitem{Niemi11}
A.~Niemi, J.~A. Bramwell, and L.~F. Demkowicz, ``Discontinuous
  {P}etrov--{G}alerkin method with optimal test functions for thin-body
  problems in solid mechanics,'' {\em Computer Methods in Applied Mechanics and
  Engineering}, vol.~200, no.~9-12, pp.~1291--1300, 2011.

\bibitem{FENICS}
M.~S. Aln{\ae}s, J.~Blechta, J.~Hake, A.~Johansson, B.~Kehlet, A.~Logg,
  C.~Richardson, J.~Ring, M.~E. Rognes, and G.~N. Wells, ``The fenics project
  version 1.5,'' {\em Archive of Numerical Software}, vol.~3, no.~100, 2015.

\bibitem{FreeFem}
F.~Hecht, ``New development in freefem++,'' {\em J. Numer. Math.}, vol.~20,
  no.~3-4, pp.~251--265, 2012.

\bibitem{GetFEM}
``{GetFEM++}.''

\bibitem{getdp}
C.~Geuzaine, ``{GetDP}: a general finite-element solver for the de {R}ham
  complex,'' in {\em PAMM Volume 7 Issue 1. Special Issue: Sixth International
  Congress on Industrial Applied Mathematics (ICIAM07) and GAMM Annual Meeting,
  Z{\"u}rich 2007}, vol.~7, pp.~1010603--1010604, Wiley, 2008.

\bibitem{hermes-adaptivity}
P.~Solin, D.~Andrs, J.~Cerveny, and M.~Simko, ``Pde-independent adaptive hp-fem
  based on hierarchic extension of finite element spaces,'' {\em J. Comput.
  Appl. Math.}, vol.~233, pp.~3086--3094, 2010.

\bibitem{hermes-hanging-nodes}
P.~Solin, J.~Cerveny, and I.~Dolezel, ``Arbitrary-level hanging nodes and
  automatic adaptivity in the hp-fem,'' {\em Math. Comput. Simul.}, vol.~77,
  pp.~117 -- 132, 2008.

\bibitem{hermes-runge-kutta}
P.~Solin and L.~Korous, ``Adaptive higher-order finite element methods for
  transient pde problems based on embedded higher-order implicit runge-kutta
  methods,'' {\em Journal of Computational Physics}, vol.~231, pp.~1635--1649,
  2012.

\bibitem{mifune2002fast}
T.~Mifune, T.~Iwashita, and M.~Shimasaki, ``A fast solver for fem analyses
  using the parallelized algebraic multigrid method,'' {\em IEEE transactions
  on magnetics}, vol.~38, no.~2, pp.~369--372, 2002.

\bibitem{karban2020fem}
P.~Karban, D.~P{\'a}nek, T.~Orosz, I.~Petr{\'a}{\v{s}}ov{\'a}, and
  I.~Dole{\v{z}}el, ``Fem based robust design optimization with agros and
  {\=a}rtap,'' {\em Computers \& Mathematics with Applications}, 2020.

\bibitem{karban2013advanced}
P.~Karban, I.~Dole{\v{z}}el, F.~Mach, and B.~Ulrych, ``Advanced adaptive
  algorithms in 2d finite element method of higher order of accuracy,'' in {\em
  Selected Topics in Nonlinear Dynamics and Theoretical Electrical
  Engineering}, pp.~255--271, Springer, 2013.

\bibitem{tsuboi1994computation}
H.~Tsuboi, F.~Kobayashi, T.~Misaki, and M.~Tanaka, ``Computation of
  three-dimensional electric field problems by a boundary integral method and
  its application to insulation design,'' {\em Periodica Polytechnica
  Electrical Engineering}, vol.~38, no.~4, pp.~381--393, 1994.

\bibitem{said1997case}
A.~R. Said, ``Case study on gis defects and new possibilities for preventive
  maintenances,'' {\em Periodica Polytechnica Electrical Engineering
  (Archives)}, vol.~41, no.~1, pp.~15--25, 1997.

\bibitem{femmimp}
D.~Meeker, ``Meshing heuristics for improved convergence (femm) {@ONLINE},''
  October 2016.

\bibitem{rudnev2017handbook}
V.~Rudnev, D.~Loveless, and R.~L. Cook, {\em Handbook of induction heating}.
\newblock CRC press, 2017.

\bibitem{artap_paper}
D.~P\'anek, T.~Orosz, and P.~Karban, ``Artap: Robust design optimization
  framework for engineering applications,'' in {\em The Third International
  Conference on Intelligent Computing in Data Sciences ICDS2019}, p.~4p, 2019.

\bibitem{induction_brazing_1}
D.~P\'anek, P.~Karban, and I.~Dole\v{z}el, ``Comparison of simplified
  techniques for solving selected coupled electroheat problems,'' {\em
  COMPEL-The international journal for computation and mathematics in
  electrical and electronic engineering}, p.~6p, 2019.

\bibitem{induction_brazing_2}
D.~P\'anek, T.~Orosz, P.~Krop\'ik, P.~Karban, and I.~Dole\v{z}el,
  ``Reduced-order model based temperature control of induction brazing
  process,'' in {\em 2019 Electric Power Quality and Supply Reliability (PQ)},
  p.~4p, 2019.

\bibitem{panek2020comparison}
D.~P{\'a}nek, P.~Karban, T.~Orosz, and I.~Dole{\v{z}}el, ``Comparison of
  simplified techniques for solving selected coupled electroheat problems,''
  {\em COMPEL-The international journal for computation and mathematics in
  electrical and electronic engineering}, 2020.

\bibitem{dolezel2010accurate}
I.~Dolezel, P.~Karban, P.~Kropik, and D.~Panek, ``Accurate control of position
  by induction heating-produced thermoelasticity,'' {\em IEEE transactions on
  magnetics}, vol.~46, no.~8, pp.~2888--2891, 2010.

\end{thebibliography}
\end{document}